\documentclass[amsmath,amssymb,superscriptaddress,longbibliography,prb,twocolumn,floatfix]{revtex4-2}

\usepackage{graphicx}
\usepackage{xcolor}
\usepackage{comment}
\usepackage[normalem]{ulem}
\usepackage{hyperref}
\hypersetup{
    colorlinks=true,
    urlcolor= blue,
    citecolor=blue,
    linkcolor= blue,
    bookmarksopen=false,
    }

\definecolor{lightgray}{gray}{0.6}
\newif\ifptitle
\newif\ifpnumber
\newcounter{para}
\newcommand\ptitle[1]{\par\refstepcounter{para}
{\ifpnumber{\noindent\textcolor{lightgray}{\textbf{\thepara}}\indent}\fi}
{\ifptitle{\textbf{[{#1}]}}\fi}}
\pnumbertrue  % comment this line to hide paragraph numbers

\newcommand{\newtext}[1]{{\textcolor{black}{#1}}}
\newcommand{\newnewtext}[1]{{\textcolor{black}{#1}}}

\newcommand{\purple}[1]{{\textcolor{black}{#1}}}
\newcommand{\Tc}{$T_c$}
\newcommand{\STO}{SrTiO$_3$}
\newcommand{\TiO}{TiO$_{2-x}$}
\newcommand{\SrO}{SrO}
\newcommand{\hphys}{Department of Physics, Harvard University, Cambridge, Massachusetts 02138, USA}
\newcommand{\heng}{School of Engineering \& Applied Sciences, Harvard University, Cambridge, Massachusetts 02138, USA}
\newcommand{\cns}{Center for Nanoscale Systems, Harvard University, Cambridge, MA, USA}

\begin{document}

\title{Imaging Se diffusion across the FeSe/SrTiO$_3$ interface}

\author{Samantha O'Sullivan}
\affiliation{\hphys}
\author{Ruizhe Kang}
\affiliation{\heng}
\author{Jules A. Gardener}
\affiliation{\cns}
\author{Austin J. Akey}
\affiliation{\cns}
\author{Christian E. Matt}
\email[]{christian.matt87@gmail.com}
\affiliation{\hphys}
\author{Jennifer E. Hoffman}
\email[]{jhoffman@physics.harvard.edu}
\affiliation{\hphys}
\affiliation{\heng}

\begin{abstract}
Monolayer FeSe on \STO\ superconducts with reported \Tc\ as high as 100 K, but the dramatic interfacial \Tc\ enhancement remains poorly understood. Oxygen vacancies in \STO\ are known to enhance the interfacial electron doping, electron-phonon coupling, and superconducting gap, but the detailed mechanism is unclear. Here we apply scanning transmission electron microscopy (STEM) and electron energy loss spectroscopy (EELS) to FeSe/\STO\ to image the diffusion of selenium into \STO\ to an unexpected depth of several unit cells, consistent with the simultaneously observed depth profile of oxygen vacancies. Our density functional theory (DFT) calculations support the crucial role of oxygen vacancies in facilitating the thermally driven Se diffusion. In contrast to excess Se in the FeSe \textit{monolayer} or FeSe/\STO\ \textit{interface} that is typically removed during post-growth annealing, the diffused Se remains in the top few unit cells of the \STO\ \textit{bulk} after the extended post-growth annealing that is necessary to achieve superconductivity. Thus, the unexpected Se in \STO\ may contribute to the interfacial electron doping and electron-phonon coupling that enhance \Tc, suggesting another important role for oxygen vacancies as facilitators of Se diffusion.
\end{abstract}

\maketitle

\section{\label{sec:Intro}Introduction}

\ptitle{High Tc in FeSe/STO interface} Monolayer FeSe grown on \STO\ (STO) superconducts with a transition temperature $T_c$ as high as 100~K \cite{WangCPL2012, HuangARCMP2017, GeJNatMat2015}, an order of magnitude higher than bulk FeSe ($T_c \sim 8.8$~K \cite{BohmerPRB2016}). 
While there is general consensus that the interface plays a crucial role in the enhanced superconductivity \cite{ZhangWPRB2014,LiuDNatComm2012, HeSNatMat2013, TanSNatMat2013, ZhangHNatComm2017, ZhaoWSciAdv2018,LeeNature2014, SongQNatComm2019, ZhangWNanoLett2016, ZhangSPRB2016}, the specific mechanism remains controversial. Angle-resolved photoemission spectroscopy (ARPES), electron energy loss spectroscopy (EELS), and scanning tunneling microscopy (STM) found evidence for a cooperative interplay of two effects: substrate-induced electron doping \cite{LiuDNatComm2012, HeSNatMat2013, TanSNatMat2013, ZhangHNatComm2017, ZhaoWSciAdv2018} and interfacial electron-phonon coupling \cite{LeeNature2014, SongQNatComm2019, ZhangWNanoLett2016, ZhangSPRB2016, GongGPRB2019}. 
But the wide range of measured \Tc\ in nominally similar samples suggests that both effects are strongly influenced by the detailed atomic structure and chemical composition of the interface.
% The wide variation of superconducting gap with interfacial oxygen vacancy and excess Se concentrations, and with atomic distances, further suggests that electron doping and electron-phonon coupling are both strongly affected by the atomic structure and chemical composition of the interface.

\ptitle{Oxygen in STO} Oxygen plays a key role in both electron doping and electron-phonon coupling at the STO interface. Oxygen vacancies directly donate charge carriers \cite{ZhangWPRB2014, HeSNatMat2013, LiuDNatComm2012, TanSNatMat2013, BangJPRB2013, LiF2DMat2016}, or indirectly alter the STO work function and associated charge transfer induced by band bending \cite{ZhaoWSciAdv2018}. 
On the other hand, STO surface oxygen and its substitutions control the energy and form of the phonon modes that couple to the FeSe electrons \cite{LeeNature2014, ZhangSPRB2016, GongGPRB2019}. Such electron-phonon coupling strongly influences \Tc\ 
% as shown by ARPES
\cite{LeeNature2014, SongQNatComm2019, ZhangHNatComm2017}, but could be screened by excess Se at the interface
% as suggested by EELS
\cite{LiF2DMat2016}. Finally, the pronounced dependence of electron-phonon coupling on oxygen vacancy concentration \cite{ChenCNatComm2015, WangZNatMat2016} complicates the interplay between the electron doping and electron-phonon coupling contributions to \Tc. The fact that enhanced superconductivity has been found in monolayer FeSe grown on various oxides, including anatase TiO$_2$ \cite{DingHPRL2016}, BaTiO$_3$ \cite{PengRNatComm2014}, LaTiO$_3$ \cite{JiaAdvSci2021}, NdGaO$_3$ \cite{YangScienceBulletin2019}, and MgO \cite{ZhaoGScienceBulletin2018} -- while absent in non-oxide systems \cite{SongCPRB2011,ZhangWNanoLett2016} -- further emphasizes the importance of oxygen chemistry on FeSe superconductivity.

\ptitle{Interface chemistry} 
% It is well known that oxygen vacancies are formed in STO during high temperature vacuum annealing \cite{HenslingFNature2017,ZvanutJAP2008}. 
Selenium belongs to the same chemical family as oxygen, which suggests that Se atoms might fill the O vacancies that typically form during high temperature vacuum annealing \cite{TennePRB2007, SzotPRL2002, MerklePCCP2003, ZvanutJAP2008}. Indeed such a scenario has been theoretically predicted for oxygen vacancies in the top \TiO\ layer \cite{BangJPRB2013} and experimentally supported by ARPES \cite{HeSNatMat2013} and scanning transmission electron microscopy (STEM) \cite{LiF2DMat2016, GongGPRB2019}. 
% However, so far no experiment has tested Se diffusion into subsurface STO layers. 
Furthermore, several groups employ high-temperature annealing under high Se pressure to prepare the STO surface prior to FeSe growth \cite{WangCPL2012,XuPhysChemChemPhys2017}, which might enhance Se diffusion into STO as more oxygen vacancies are created and the formation energy for Se substitution is lowered. Often, excess Se in the FeSe film and at the interface is removed during post-growth annealing \cite{LiF2DMat2016}, which might not be possible for Se diffused deeper into the STO subsurface. 
Although accurate knowledge of the interface chemical composition is of profound importance for exact modeling of the superconductivity enhancement in the FeSe/STO heterostructure, no experiment has investigated Se diffusion into subsurface layers of STO.

\begin{figure}
	\includegraphics[width=\columnwidth]{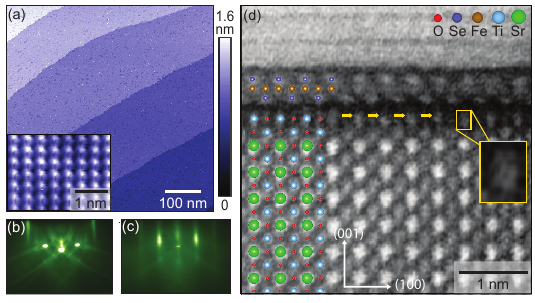}
 	\caption{High-quality monolayer FeSe grown on \STO\ (STO). (a) Topography acquired by STM, with inset showing atomically resolved top Se layer (sample bias $V_\mathrm{s} = 0.1$~V, current setpoint $I_\mathrm{s} = 90$~pA). (b-c) RHEED images of STO substrate along (100) before and after FeSe film growth (recorded at 15 keV, $T = 400$ $^\circ$C). (d) Atomic resolution STEM image of monolayer FeSe on STO substrate with Te capping layer. Yellow arrows and zoom-in box highlight elongated shape of the top Ti atoms. FeSe/STO crystal structure is overlaid on left side of (d).}
	\label{fig:fig1}
 \end{figure}
 
\begin{figure*}
  	\includegraphics[width=\textwidth]{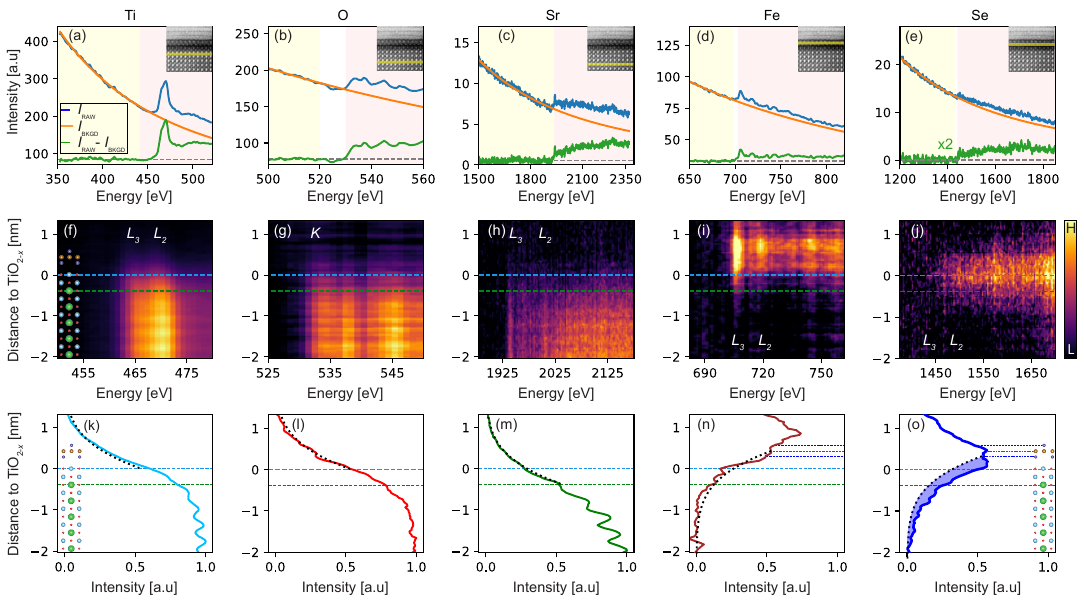}
  	\caption{(a)-(e) EELS signals and the background subtraction process for elements Ti, O, Sr, Fe, and Se. The background fitting range is shaded yellow and the signal integration range is shaded pink, with the raw signal intensity ($I_{\mathrm {RAW}}$, \newnewtext{averaged along the yellow line in the corresponding inset}) in blue, fit background intensity ($I_{\mathrm {BKGD}}$) in orange, and difference of the two ($I_{\mathrm {RAW}}$ - $I_{\mathrm {BKGD}}$) in green. Arbitrary vertical offset (gray dashed line) has been added to each ($I_{\mathrm {RAW}}$ - $I_{\mathrm {BKGD}}$) curve for visual purposes. (f)-(j) Energy loss within a single measurement region, averaged along the (100) direction at indicated absorption edges after background subtraction ($I_{\mathrm {RAW}}$ - $I_{\mathrm {BKGD}}$), with the color bar denoting low (L) to high (H) intensity. Dashed blue (green) lines indicate positions of the upper \TiO\ (\SrO) layers. (k-o) Energy-integrated linecuts of energy loss plots averaged over six measurement regions. Maximum intensity of Ti, O, Sr are normalized to 1. Fe and Se are normalized to the expected intensity of a single layer using the resolution broadening (black dotted line) inferred from the Sr edge (see Fig.\ \ref{fig:figS1}). Resolution-broadened beam profile is shown on each panel as an identical black dotted curve, shifted or reflected appropriately.}
 	\label{fig:fig2}
  \end{figure*}

\begin{figure}[t!]
	\includegraphics[width=\columnwidth]{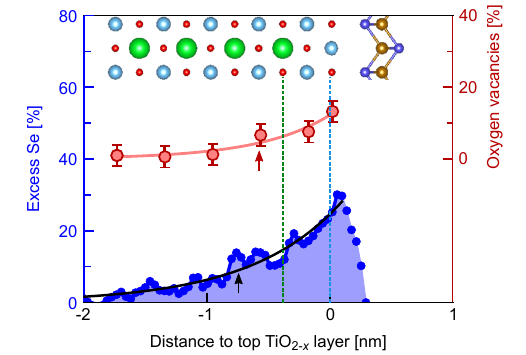}
 	\caption{Excess selenium at substrate surface correlates with oxygen vacancy formation. Left axis: Excess selenium (as a fraction of the lower Se layer in FeSe) calculated by subtracting the dotted black line from the Se signal in Fig.\ \ref{fig:fig2}(o). Right axis: Increasing formation of oxygen vacancies towards the STO surface extracted from O$_2$/Ti ratio of linecuts in Fig.\ \ref{fig:fig2}(k)-\ref{fig:fig2}(l). Vertical arrows indicate decay length.}
	\label{fig:fig3}
 \end{figure}
 
\begin{figure}[t!]
	\includegraphics[width=\columnwidth]{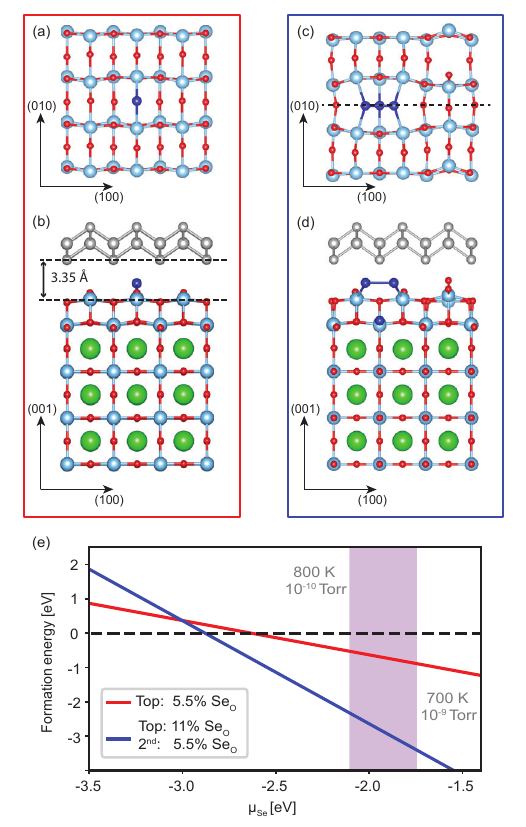}
 	\caption{(a)-(b) Top view and side view of fully relaxed structure, where a single Se atom (dark blue sphere) substitutes one oxygen vacancy on the top \TiO\ layer, corresponding to 5.5\% Se$_\mathrm{O}$ substitution. (c)-(d) Same as (a-b) but with three Se atoms substituting two oxygen vacancies (11\% Se$_\mathrm{O}$) in the top \TiO\ layer and one oxygen vacancy (5.5\% Se$_\mathrm{O}$) in the second TiO$_{2-y}$ layer, at locations as indicated. Note that only the top two layers of atoms are shown in (a) and (c). The FeSe layers in (b) and (d) are shown in gray as a guide to the eye and are not included in the DFT calculation. (e) Formation energy as a function of the Se chemical potential. The purple region indicates realistic experimental conditions with substrate temperatures and Se partial pressures as indicated.}
	\label{fig:fig4}
 \end{figure}
  
\ptitle{Here we show...} Here we use STEM and EELS to reveal the diffusion of Se several unit cells deep into STO % (Figs.\~\ref{fig:fig1},\ref{fig:fig2}) 
that occurs during the monolayer FeSe film growth and annealing, both performed at temperatures below $\sim520^\circ$C. We find that the excess Se decays exponentially into STO, as predicted by Fick's law of thermally activated elemental diffusion \cite{kasap2017}. Furthermore, we observe a similar line profile and decay length of oxygen vacancies at the STO surface % (Fig.\ \ref{fig:fig3}) 
which, in combination with density functional theory (DFT) calculations, suggests that oxygen vacancies play a pivotal role for Se diffusion. % (Fig.\~\ref{fig:fig4}). 
The role of oxygen vacancies in facilitating Se diffusion is further supported by the contrast between the Se and Fe line profiles and the negligible diffusion of Fe, which belongs to a different chemical family and therefore does not substitute for oxygen. 

\section{\label{sec:Methods}Methods}
\ptitle{Growth methods} Monolayer FeSe was grown by molecular beam epitaxy (MBE) on a Nb-doped (0.05\%) STO(001) substrate from Crystek. The STO substrate was etched with buffered HF (NH$_{4}$F\,:\,HF~=~7\,:\,1, diluted with equal volume of deionized water) for 30 seconds, then annealed in O$_{2}$ at 950 $^{\circ}$C for 1 hour. The substrate was transferred into the MBE chamber (base pressure $<5 \times 10^{-10}$ Torr) and degassed for 3 h at 500$^{\circ}$C. Importantly, \textit{no} high-temperature Se molecular beam etching was performed prior to growth \cite{WangCPL2012}. FeSe was deposited in three rounds by co-evaporating Fe (99.995\%) and Se (99.999\%) with a molar flux ratio of 1:30 and substrate temperatures between 400$^{\circ}$C and 520$^{\circ}$C, followed by post-growth annealing at 450$^{\circ}$C~-~520$^{\circ}$C 
% \deltext{(see Supplementary Information \cite{SI} for detailed recipe). }
\newtext{(First round:  0.95 unit cells FeSe deposited at a substrate temperature of $400^{\circ}$C and 3~h post-growth annealing at $450^{\circ}$C. 
Second round: 0.3 unit cells FeSe at $400^{\circ}$C with 3~h post-growth annealing at $450^{\circ}$C and 4~h at $520^{\circ}$C. 
Third round: $\sim0.2$ unit cells FeSe at $520^{\circ}$C, post-growth annealed at the same temperature for 4~h.)} 
The final annealing step in ultrahigh vacuum (UHV, $<5 \times 10^{-10}$ Torr) was performed at $\sim 510^{\circ}$C for 10~h. After each growth step the sample was transferred through UHV to a home-built scanning tunneling microscope (STM) for imaging at $\sim77$~K. \newtext{The final STM scan confirmed that the high annealing temperature $\gtrsim 500^{\circ}$C was effective in removing all 2-unit-cell islands \cite{ZhangHNatComm2017}.} Finally, the film was capped with a $\sim 40$~nm Te layer \cite{LiPRM2021} at room temperature to prepare for cross-sectional STEM and EELS measurements. A lamella of thickness $30\pm6$ nm was prepared using focused ion beam milling (FEI Helios 660). A JEOL ARM 200F operated at 200~kV was used to record room temperature STEM (JEOL HAADF detector) and EELS measurements at six different locations of the lamella. EELS data was acquired with STEM probe settings of 197~pA current and 22.4~mrad convergence angle, using a Gatan Enfinium EELS spectrometer. \newtext{We grew a second sample for low-temperature STM imaging, and confirmed a superconducting gap of $\sim 15$ meV at $T = 4.7$ K.}

We performed DFT calculations using the open-source Quantum Espresso (QE) software package \cite{HenslingFNature2017,ZvanutJAP2008}. We constructed a $3 \times 3\times 3$ STO supercell, terminated by double \TiO\ layer and added $20\; \mathrm{\AA}$ of vacuum spacing along the (001) axis to simulate the two dimensional surface structure using periodic boundary conditions. We used ultra-soft pseudo-potentials for Sr, Ti and Se atoms and projector augmented-wave pseudo-potential for O atoms. We set the kinetic energy cutoff to be 40 Ry and the charge density cutoff to 400 Ry. We used a Gaussian smearing of 0.01~Ry to improve the convergence during the relaxation. We relaxed the entire structure until both the forces and total energy for ionic minimization was smaller than $1 \times 10^{-4}$ Hartree/Bohr and $1 \times 10^{-4}$ Hartree, respectively. The energy convergence threshold for self-consistency was $1 \times 10^{-6}$ Hartree. We sampled the first Brillouin zone by a $4 \times 4 \times 1$ $k$-grid. When relaxing the structure, we allowed only the top three atomic layers (top \TiO, second TiO$_{2-y}$, top SrO) to move, while the rest was fixed.

\section{\label{sec:Results}Results}

\ptitle{Film characterization -- Fig1} 
The high crystalline quality of the FeSe film is apparent in the STM topography in Fig.\ \ref{fig:fig1}(a) showing a uniform monolayer coverage and atomically smooth surface areas (see inset). Reflection high-energy electron diffraction (RHEED) images of the STO surface in Fig.\ \ref{fig:fig1}(b) show sharp diffraction spots, indicating a non-reconstructed ($1\times1$) termination. The post-growth RHEED image in Fig.\ \ref{fig:fig1}(c) depicts the typical pattern for epitaxial monolayer FeSe \cite{HuangARCMP2017}. Figure\ \ref{fig:fig1}(d) shows our atomic resolution cross-sectional STEM measurement in which we can identify the Te capping layer, the monolayer FeSe, and an atomically sharp FeSe/STO interface.
% We overlay the crystal structure of FeSe/STO to indicate the position of the Fe, Se, Sr, O, and Ti atoms which allow for careful analysis of the interfacial atomic structure. 
We measure the inter-atomic distance between the bottom Se and top Ti layers to be $3.35 \pm 0.21$ \AA, consistent with previous STEM measurement \cite{LiF2DMat2016}. We also observe the double layer \TiO\ termination of the STO, which is commonly seen \cite{LiF2DMat2016, ErdmanNNature2002, ZhuGNature2012, MarshallMSpringer2015, ZhaoWSciAdv2018}. 
% We will refer to the \TiO layer higher in the (001) direction as the ``top \TiO'', and the layer lower in the (001) direction as the ``bottom \TiO''. 
While we don't observe any ordered Se layer between the FeSe and the top \TiO\ layer \cite{ZhaoWSciAdv2018} we note that the top Ti atoms appear slightly elongated along the (001) direction, which has been interpreted as a sign of additional Se  at the interface \cite{LiF2DMat2016}.

\ptitle{EELS background \& Se -- Fig.\ 2} To identify the chemical composition of the interface, we analyze the EELS measurement over a wide energy range covering Ti, Fe, Se, O, and Sr absorption edges. We average the absorption spectra along the (100) direction of the scan window and subtract a power law background, as shown in Figs.\ \ref{fig:fig2}(a-e) (see also Appendix \ref{app:SeProfile} and Fig.\ \ref{fig:figSeEdge}).
Figures\ \ref{fig:fig2}(f-j) show a resolution-limited cutoff for Ti, O, and Sr above the top \TiO\ layer and for Fe below the \TiO\ layer, as expected for an atomically sharp interface. % Surprisingly, we find that the Se intensity is centered around the top \TiO\ layer (Fig.\ \ref{fig:fig2}j), suggesting that Se diffused into the STO substrate. 
In contrast, we find that the Se intensity has a longer tail below the top \TiO\ layer shown in Fig.\ \ref{fig:fig2}(j), suggesting that Se diffused into the STO substrate. 
This observation is confirmed in the energy-integrated linecuts, shown in Figs.\ \ref{fig:fig2}(k-o). The intensity drop of Ti, O above the topmost \TiO\ layer and Sr above the SrO layer are determined by the beam shape of the STEM probe (see Appendix \ref{app:STEMres} and Fig.\ \ref{fig:figS1}). The Fe linecut follows the same expected resolution-limited intensity profile, dropping just above the \TiO\ line. However, the Se linecut deviates significantly from the expected profile and extends at higher intensities for several STO subsurface layers, indicating a significant concentration of Se below the top \TiO\ layer. 

\ptitle{Compare Se and Fe diffusion} The contrast between Se and Fe \purple{\textit{downward}} diffusion is shown by the differing deviations of their measured linecuts from their expected resolution-broadened intensity profiles, in Figs.\ \ref{fig:fig2}(n-o). While the excess Fe signal below the Fe layer in Fig.\ \ref{fig:fig2}(n) is within the instrument broadening and noise level, the excess Se, marked by the blue shaded area in Fig.\ \ref{fig:fig2}(o), is significant and extends deep into the subsurface layer of STO. \purple{The Fe intensity peak \textit{above} the FeSe monolayer may indicate the presence of excess Fe that formed FeTe islands during the Te capping process, as previously suggested by Refs.\ \cite{HuYPRB2018, HuPRB2020}.}

\ptitle{Se and O vacancies -- Fig.\ 3} To investigate the origin of Se diffusion into STO and a possible connection with preformed O vacancies, we analyze their spatial profile across the interface in Fig.\ \ref{fig:fig3}. The excess Se signal peaks just above the \TiO\ layer and falls exponentially along the (00$\bar{1}$) direction with decay length $\xi_\mathrm{Se} = 0.74 \pm 0.05$ nm (vertical black arrow). The peak position above the \TiO\ layer demonstrates excess Se between STO and the FeSe layer, consistent with  Fig.\ \ref{fig:fig1}(d) and previous STEM studies \cite{ZhaoWSciAdv2018, LiF2DMat2016}. The exponential profile is a solution of Fick's diffusion law \cite{kasap2017}, which points towards thermally-driven diffusion along an element concentration gradient, as is often observed at interfaces \cite{casey1975}. 
However, the contrasting absence of Fe diffusion into the STO suggests that thermal activation alone is not sufficient, and a second mechanism must contribute to the Se diffusion into the STO substrate. Selenium belongs to the same chemical family as oxygen, suggesting that oxygen vacancies may be partially filled with Se, similar to predictions for the top \TiO\ layer \cite{BangJPRB2013,LiF2DMat2016}.
Fig.\ \ref{fig:fig3} shows the concentration of O vacancies extracted from the spatial dependence of the O$_2$/Ti ratio of the EELS linecuts in Figs.\ \ref{fig:fig2}(k-l). We find an exponential decay length of $\xi_\mathrm{O_\mathrm{vac}} = 0.57 \pm 0.30$ nm (red arrow), corresponding to an O vacancy concentration of $11\pm 3\%$ for the top \TiO\ layer and $6\pm 3\%$ for the second TiO$_{2-y}$ layer. The consistency between $\xi_\mathrm{Se}$ and $\xi_\mathrm{O_\mathrm{vac}}$ supports the hypothesis that oxygen vacancies are crucial in facilitating the Se diffusion into STO.

%Although the decay lengths are consistent within uncertainty, the slightly longer Se decay length may indicate that thermally driven diffusion due to the Se concentration gradient (Fick's law) is acting in addition to the oxygen vacancies. 

% The slightly longer decay length of the excess Se concentration ($\xi_\mathrm{Se} = 0.74 \pm 0.05$ nm vs. $\xi_\mathrm{O_\mathrm{vac}} = 0.57 \pm 0.30$ nm, vertical arrows), is an indication that the thermally driven diffusion due to the concentration gradient (Fick's law) is acting in addition to the oxygen vacancies in the case of Se.

% O/Ti: -0.48 nm, L2/L3: -0.44

\ptitle{DFT -- Fig.\ 4} To further investigate the role of oxygen vacancies on Se diffusing into the STO surface, we use DFT to calculate the formation energies for various vacancy configurations (for more details, see Methods and Appendix \ref{app:SeDiffEnergy}). In our calculations we assume that oxygen vacancies form during vacuum annealing prior to FeSe growth. We investigate the following two configurations:
$(i)$ One vacancy per supercell in the top \TiO\ layer, corresponding to 5.5\% O vacancies,  and $(ii)$ two vacancies (11\%) in the top \TiO\ layer and one vacancy (5.5\%) in the second TiO$_{2-y}$ layer, corresponding to the measured amount of oxygen vacancies in Fig.\ \ref{fig:fig3} (see Appendix \ref{app:SeDiffEnergy} and Fig.\ \ref{fig:figSDFT}). Figures\ \ref{fig:fig4}(a-d) show the final relaxed structures after we filled each oxygen vacancy with Se (Se$_\mathrm{O}$). Our calculation shows that Se$_\mathrm{O}$ in the top \TiO\ layer protrudes slightly from the layer, consistent with our experimental observation of apparent Ti atom elongation in Fig.\ \ref{fig:fig1}(d), and the excess Se peak just above the \TiO\ layer in Fig.\ \ref{fig:fig3}. \newtext{In contrast, the Se$_\mathrm{O}$ in the second TiO$_{2-y}$ layer remains at its initial location, suggesting that diffused Se predominantly occupies the oxygen vacancy sites instead of interstitial locations.}
We next calculate the formation energies for Se atoms filling the vacancies,
\begin{equation} \label{eqn:1}
E_\mathrm{form} = E^\mathrm{DFT}_{n\mathrm{Se}} -E^\mathrm{DFT}_{n\mathrm{O_{vac}}} - n \mu_\mathrm{Se}.
\end{equation}
Here, $E^\mathrm{DFT}_{n\mathrm{O_{vac}}}$ and $E^\mathrm{DFT}_{n\mathrm{Se}}$ are the energies of the fully relaxed structure with $n$ oxygen vacancies and $n$ Se substitutions, respectively, and $\mu_\mathrm{Se}$ is the temperature and pressure dependent chemical potential of a single Se atom.  In Fig.\ \ref{fig:fig4}(e), we find that $E_\mathrm{form}$ is negative for a range of $\mu_\mathrm{Se}$  corresponding to experimental substrate temperatures and Se partial pressures, marked by the purple shaded area (see Appendix \ref{app:SeDiffEnergy} and \ref{app:SeChemPot}). Our calculations thus suggest that Se diffusion below the top \TiO\ layer is energetically favorable in presence of oxygen vacancies.

% \ptitle{DFT/Fig4, Ruizhe's version} To further understand the fact that selenium atoms tend to diffuse into the substrate surface, we carry out DFT calculations to reveal the possible configurations and their formation energy. We construct a $3 \times 3\times 3$ STO supercell with an extra \TiO layer on top. An additional 20 $\AA$ of vacuum spacing along the c axis is added to simulate the two dimensional surface structures using periodic boundary condition (for more calculation details, see Supplementary Information \cite{SI}). Optimal surface structures are found by fully relaxing the top three atomic layers and keeping the rest fixed. Since Fig.\ref{fig:fig4}(a-b) shows the 
 
\ptitle{Discuss effect of extra Se} We consider the implications of Se diffusion for the charge carrier concentration in STO.
% substrate-induced electron doping of the monolayer FeSe. 
While O vacancies create free electrons at the STO surface, which likely dope the monolayer FeSe \cite{ZhangWPRB2014, HeSNatMat2013}, excess Se has been theoretically \cite{BerlijnPRB2014, ShanavasPRB2015} and experimentally \cite{HeSNatMat2013, ZhangWPRB2014} shown to act as hole dopants.
%\orange{In other words, excess Se might increase the STO work function, compared to STO with unfilled O vacancies, which will decrease  the electron transfer into FeSe \cite{ZhaoWSciAdv2018}}. 
However, as the electronegativity of Se (2.55) is lower than that of O (3.44) we expect that even in the extreme case of all O vacancies being filled with Se, there will remain excess free electrons.
Furthermore, the excess Se could influence the STO work function and the associated interfacial band bending, altering the electron transfer into FeSe \cite{ZhaoWSciAdv2018}. The STO charge carrier concentration also modifies the electron-phonon coupling \cite{ChenCNatComm2015, WangZNatMat2016}. Additional theoretical and experimental study is required to understand the detailed effects of subsurface Se on the interfacial electron-phonon coupling, charge transfer and superconductivity enhancement.

\section{\label{sec:Conclusion}Conclusion}
\ptitle{Conclusion} To conclude, we imaged the monolayer FeSe/STO interface using atomic-resolution STEM and EELS, and we observed Se diffusion several unit cells deep into the STO. Our EELS measurements further revealed oxygen vacancies in the surface and subsurface layers of STO which, in combination with our DFT calculations, supports the scenario that oxygen vacancies are crucial to facilitate the Se diffusion. Surprisingly, the diffused Se persisted in the STO even after extended ($\sim 10$~h) post-growth UHV annealing above 500$^{\circ}$C, which has been shown to remove excess Se from the FeSe layer and the immediate interface between FeSe and STO \cite{LiF2DMat2016}. The post-growth anneal is a crucial step to obtain the high-temperature superconductivity in the FeSe/STO heterostructure \cite{HeSNatMat2013, LiF2DMat2016}. Our findings call for \newtext{future experiments to measure the relation between Se diffusion depth and superconducting \Tc, and future theoretical models to calculate the effects of Se diffusion on electron-phonon coupling and interfacial doping.} Our observation may also help to resolve the inconsistency between the calculated \cite{tikhonova_effect_2020} and experimentally measured band structure of the monolayer FeSe/STO heterostructure \cite{LiuDNatComm2012, LeeNature2014}.
 \vspace{1cm}

% \cite{Liu_DFT_PRB, Zhou_DFT_PRB, Huang_DFT_PRB, tikhonova_effect_2020, BangJPRB2013, Cao_DFT_PRB}

 \section*{Acknowledgements}
We thank Yu Xie, Boris Kozinsky, Dennis Huang, and Jason Hoffman for insightful discussions. 
% Experimental work was supported by the Center for the Advancement of Topological Semimetals (CATS), an Energy Frontier Research Center funded by the U.S. Department of Energy (DOE), Office of Science, Basic Energy Sciences (BES) through the Ames Laboratory under its Contract No. DE-AC02-07CH11358.
This work was performed in part at the Center for Nanoscale Systems (CNS), a member of the National Nanotechnology Coordinated Infrastructure Network (NNCI), which is supported by the National Science Foundation under NSF award no. 1541959. CNS is part of Harvard University.
C.E.M. was supported by the Swiss National Science Foundation under fellowships P2EZP2\_175155 and P400P2\_183890, and by the Office of Naval Research grant N00014-18-1-2691.

%%%%%%%%%%%%%%%%%%%%%%%%%%%%%%%%%%%%%%%%%%%%%%%%%%%%%%%%%%%%%%%%%%%%%%%%
%%%%%%%%%%%%%%%%%%%%%%%  Appendix %%%%%%%%%%%%%%%%%%%%%%%%
%%%%%%%%%%%%%%%%%%%%%%%%%%%%%%%%%%%%%%%%%%%%%%%%%%%%%%%%%%%%%%%%%%%%%%%%

\appendix
\section{\label{app:SeProfile}Selenium absorption edge}

\begin{figure}[h!]
	\includegraphics[width=1\columnwidth]{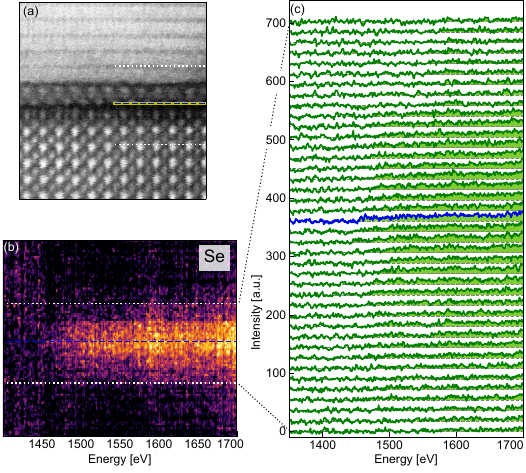}
 	\caption{Spatial evolution of Se absorption edge. 
  (a) High-resolution STEM image as shown in Fig.\ \ref{fig:fig1}(d). (b) \newnewtext{Background-subtracted} energy loss spectra close to the Se $L$ absorption edge, averaged along the (100) direction of (a), as shown in Fig.\ \ref{fig:fig2}(j). (c) Line plots of \newnewtext{background-subtracted} Se energy loss spectra for (001) range between white dotted lines in (a) and (b). \newnewtext{Blue lines in all three panels indicate the position of the lower Se layer, where the spectrum presented in Fig.\ \ref{fig:fig2}(e) was extracted.}}
	\label{fig:figSeEdge}
 \end{figure}
 
Figure \ref{fig:figSeEdge} shows the evolution of the Se absorption spectra across the FeSe/STO interface, \newnewtext{within one representative region}. The location of the lower Se layer is marked by a blue line in all three panels. The Se $L$ edge after background subtraction is shown in false color in Fig.\ \ref{fig:figSeEdge}(b) and as a linecut in Fig.\ \ref{fig:figSeEdge}(c). The total Se signal for each position along the (001) direction is the integrated green area under each curve. We reproduced this data in six distinct regions along the FeSe/STO lamella, and we show the averaged results in Fig.\ \ref{fig:fig2}(o). 

\section{\label{app:STEMres}STEM spatial resolution}
\begin{figure}[ht]
	\includegraphics[width=1\columnwidth]{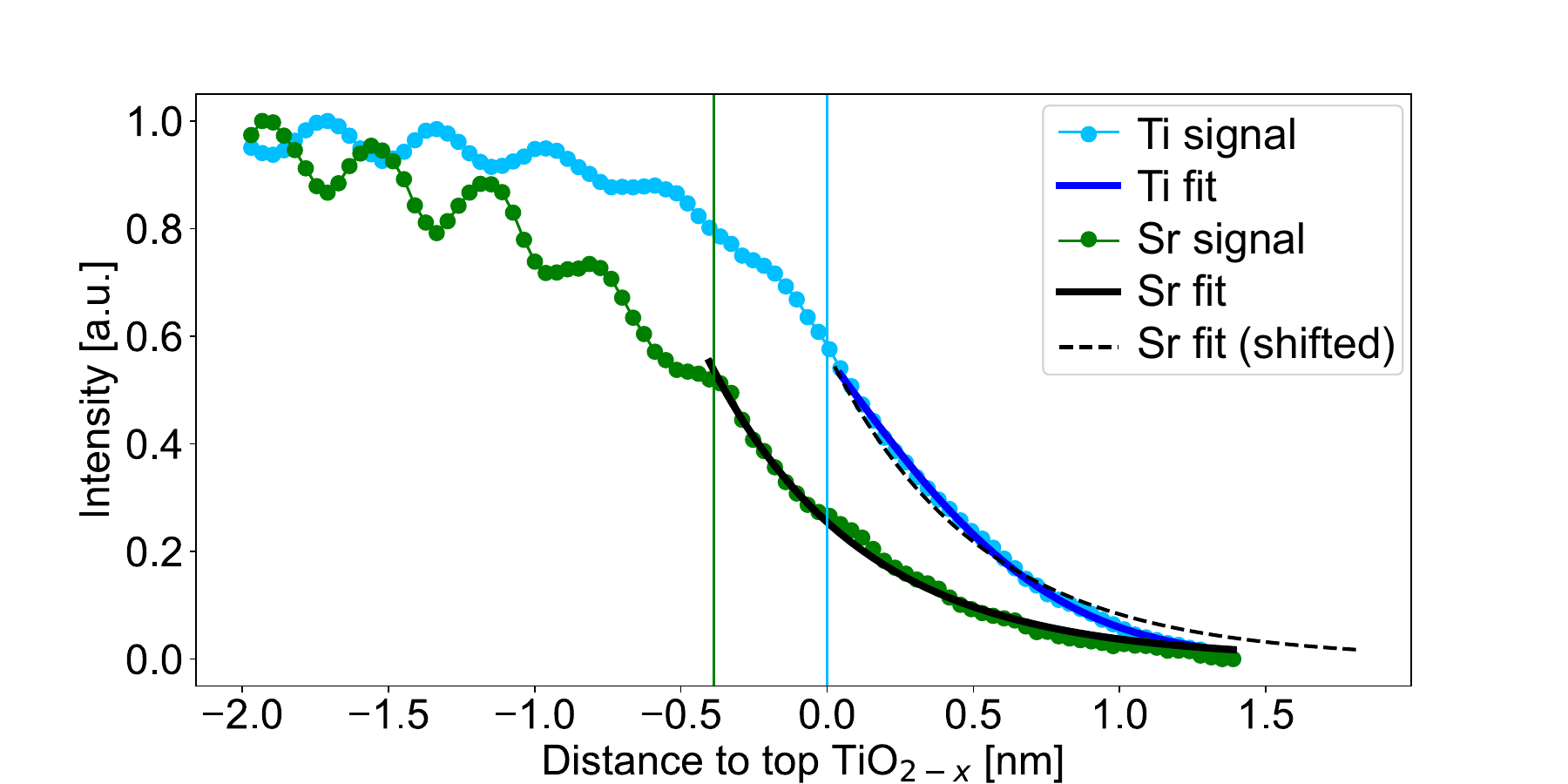}
 	\caption{Instrumental resolution. 
 	Tails of titanium (blue) and strontium (green) lineshapes of Fig.\ \ref{fig:fig2}(k,m) fit by Gaussian curve (solid lines) above the top Ti and Sr layers in STO, respectively. Location of top Ti and Sr layers are indicated by blue and green vertical lines, respectively. We assume a stoichiometric Sr layer and no Sr above, thus the Sr fit (black solid line) determines the broadening of the STEM signal due to the finite width of the electron beam.}
	\label{fig:figS1}
 \end{figure}
 To determine spatial broadening due to the finite STEM electron beam width, we fit Gaussian curves to the tails of the Ti and Sr EELS signals that extend above the top Ti and Sr layers in Figs.\ \ref{fig:fig2}(k,m). In Fig.\ \ref{fig:figS1}, we show both signals and fits, and we also shift the Sr fit to overlap with the Ti fit, which demonstrates a similar lineshape (Sr is slightly broader). The similarity indicates a spatial resolution that is almost independent of energy (absorption energies: Ti $L_3$ edge: 456~eV, Sr $L_3$ edge: 1940~eV). We then compare the Sr fit tail (representing pure instrument broadening) to our Fe and Se line profiles in Figs.\ \ref{fig:fig2}(n) and \ref{fig:fig2}(o) to determine the Fe and Se excess signal that corresponds to real element diffusion into the STO bulk. Since the monolayer FeSe consists of only one Fe layer and two Se layers, we expect the peak amplitude of these element profiles to be reduced due to the finite spatial resolution. We therefore normalize the Fe and Se profiles in Fig.\ \ref{fig:fig2}(n,o) such that the intensity of the Fe (Se) profile at the location of the Fe (Se) layer equals the Sr signal at the top Sr layer (which is less than the Sr signal in the bulk of STO).

\section{\label{app:SeDiffEnergy}Formation energy of selenium diffusion}

\begin{figure}
	\includegraphics[width=1\columnwidth]{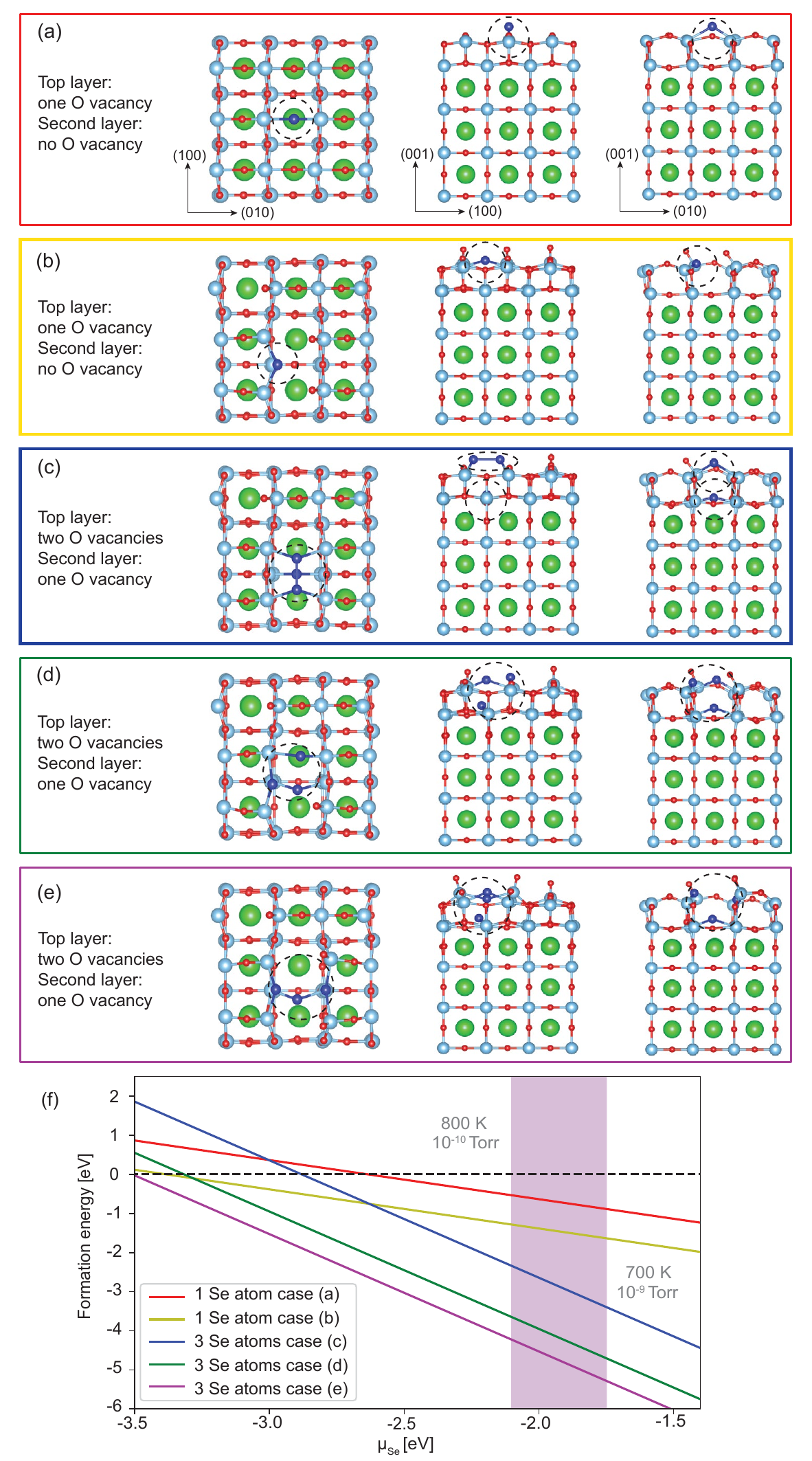}  
 	\caption{Relaxed crystal structures and formation energy of Se-substituted oxygen vacancies.  (a-e) 
 	Relaxed $3 \times 3 \times 3$ supercell with (a-b) one oxygen vacancy replaced by Se$_\mathrm{O}$ in the top \TiO\ layer ($2x \sim 5.5\%$) and (c-e) three oxygen vacancies replaced by Se$_\mathrm{O}$, two in the top \TiO\ layer ($2x = 11\%$) and one in the second TiO$_{2-y}$ layer ($2y = 5.5\%$), calculated for three different configurations. In each case, Se in the top \TiO\ layer is slightly protruding out-of-plane giving rise to the vertically elongated Ti appearance in the STEM images of Fig.\ \ref{fig:fig1}(d) and reported in Ref.\ \cite{LiF2DMat2016}.  (f) Formation  energies for cases (a-e), calculated using Eq.\ \ref{eq:1}, demonstrating that it is favorable (lowers the energy) for the system to fill in  preformed oxygen vacancies with Se atoms under realistic experimental conditions of Se partial pressure and sample temperature (purple shaded area). 
  }
	\label{fig:figSDFT}
 \end{figure}

Electrical \cite{SzotPRL2002}, magnetic resonance \cite{ZvanutJAP2008, MerklePCCP2003}, and optical studies \cite{TennePRB2007} have shown that oxygen vacancies occur during heat treatment in vacuum  near the surface of STO substrates. Thus our DFT calculations start from supercell models that have oxygen vacancies on the top two layers of \TiO.
We define the formation energy using
\begin{equation} \label{eq:1}
E_\mathrm{form} = E^\mathrm{DFT}_{n\mathrm{Se}} - E^\mathrm{DFT}_{n\mathrm{O_{vac}}} - n \mu_\mathrm{Se} 
\end{equation}
\noindent where $E^\mathrm{DFT}_{n \mathrm{Se}}$ is the energy of the fully relaxed structure with Se implemented in either the top or the second \TiO\ layer. $E^\mathrm{DFT}_{n\mathrm{O_{vac}}}$ is the energy of the fully relaxed STO supercell with vacancies in either the top or the second \TiO\ layer. Both of them are calculated by DFT. $\mu_\mathrm{Se}$ is the chemical potential of a single Se atom, which is a function of temperature $T$ and pressure $p$. $\mu_\mathrm{Se}$ can be written as
\begin{equation} \label{eq:2}
 \mu_\mathrm{Se} (T, p) = \frac{1}{2} \mu_\mathrm{Se_{2}} = \frac{1}{2} (E_\mathrm{Se_2}^\mathrm{DFT} + \mu_\mathrm{Se_{2}}(T, p))
\end{equation}
\noindent where $E_\mathrm{Se_2}^\mathrm{DFT}$ is the energy of an isolated Se dimer molecule as calculated by DFT. Since it is well known that DFT tends to overbind the molecule \cite{PhysRevB.91.245114}, we then use Eq.\ \ref{eq:2add} to finally determine the energy of the Se$_2$ molecule
\begin{equation} \label{eq:2add}
 E_\mathrm{Se_2}^\mathrm{DFT} = 2 E_\mathrm{Se}^\mathrm{DFT} - E_\mathrm{bond}
\end{equation}
\noindent where $E_\mathrm{Se}^\mathrm{DFT}$ is the energy of an isolated single Se atom determined by a self-consistent DFT calculation. $E_\mathrm{bond}$ is the bond energy of the $\mathrm{Se_2}$ molecules obtained from Ref.\ \cite{Inorganicchem}, and $\mu_\mathrm{Se_{2}}(T, p)$ in Eq.\ \ref{eq:2} is the chemical potential for the selenium dimer molecule, which depends on temperature and pressure, as derived in Appendix \ref{app:SeChemPot}.

To calculate the formation energy, we first fully relaxed the pristine STO supercell with a double-layer \TiO\ termination. 
% layers as is shown in Fig \ref{fig:S2}. 
The calculated distance between the double \TiO\ layers is 2.19 \AA, which is very close to our experimental value of $\sim 1.9\pm0.3$ \AA, and confirms the validity of our relaxed structure. 
We then calculated various oxygen vacancy configurations and their relaxed crystal structures and energies $E^\mathrm{DFT}_{n\mathrm{O_{vac}}}$ for each case. Consecutively we replaced each oxygen vacancy with a selenium atom and again relaxed the supercell to obtain $E^\mathrm{DFT}_{n \mathrm{Se}}$ and the final structures presented in Figs.\ \ref{fig:fig4}(a-d) and Fig.\ \ref{fig:figSDFT}.

\section{\label{app:SeChemPot}Estimation of the chemical potential of a single selenium atom}

To estimate the chemical potential of a single Se atom, we determine the chemical potential for a selenium dimer molecule. Considering the Se$_{2}$ molecule as ideal diatomic gas, its partition function has contributions from translation, vibration and rotation, which can be written as
\begin{equation}\label{eq:3}
Z = Z_\mathrm{trans}Z_\mathrm{vib}Z_\mathrm{rot}
\end{equation}

\noindent Here we ignore the contribution from the electronic levels since they will contribute to the thermodynamic properties only at high temperature or if unpaired electrons are present \cite{tiwari_thermodynamical_2012}.

Using the rigid rotor-harmonic oscillator approximation \cite{tiwari_thermodynamical_2012, stat-mech}, one can explicitly evaluate all the thermodynamic quantities. The chemical potential can be expressed in terms of a reference pressure as shown in \cite{otgonbaatar_effect_2014}
\begin{equation}\label{eq:4}
\mu_\mathrm{Se_{2}} (T, p) = \mu^{0}_\mathrm{Se_{2}} (T, p^{0}) + k_\mathrm{B}T \mathrm{log}\left(\frac{p}{p^{0}}\right)
\end{equation}

\noindent where $\mu^{0}_\mathrm{Se_{2}}$ is the chemical potential at reference pressure $p^{0}$, which is usually taken as 1 atm; $k_{B}$ is the Boltzmann constant.

\begin{table}[ht!]
\caption{\label{table:1} Calculated thermodynamical properties of selenium molecule at 1 atm.}
\centering
\begin{tabular}{c c c c} 
 $T$ (K) & $H$ (kJ/mol)\cite{tiwari_thermodynamical_2012} & $S$ (J/mol/K)\cite{tiwari_thermodynamical_2012} & $\mu^{0}_\mathrm{Se_{2}}$ (eV) \\
 \hline
 500 & 14.53 & 270.86 & -1.25 \\
 600 & 17.46 & 277.57 & -1.54 \\
 700 & 20.40 & 283.27 & -1.84 \\
 800 & 23.34 & 288.23 & -2.15 \\
 900 & 26.29 & 292.61 & -2.46 \\
 \hline
\end{tabular}
\end{table}
 
Table \ref{table:1} shows some of the calculated values of thermodynamical quantities within the temperature range that is close to our experimental condition. The enthalpy $H$ and entropy $S$ were adapted from Ref.\ \cite{tiwari_thermodynamical_2012}, and the chemical potential $\mu^{0}_\mathrm{Se_{2}}$ is defined as Gibbs free energy ($G$) per molecule,
\begin{equation}\label{eq:5}
\mu^{0}_\mathrm{Se_{2}} = \frac{G}{N_{A}} = \frac{H - TS}{N_{A}}
\end{equation}

If we insert the values from Table \ref{table:1} 
into Eq.\ \ref{eq:4}, we will obtain the chemical potential for a single Se$_{2}$ molecule at any given temperature and pressure. Given our experimental conditions, here we consider two extreme cases:

(1) For $T = 800$ K and pressure $p = 10^{-10}$ Torr, $\mu_\mathrm{Se_{2}} = -4.20$ eV, which is the lower limit.

(2) For $T = 700$ K and pressure $p = 10^{-9}$ Torr, $\mu_\mathrm{Se_{2}} = -3.50$ eV, which is the upper limit.

\noindent Plugging these values into Eq.\ \ref{eq:2}, we can estimate that under our experimental conditions, the range of the chemical potential for a single Se atom (for simplicity, here we set $E^\mathrm{DFT}_\mathrm{Se} = 0$) is
\begin{equation}
 \mu_\mathrm{Se} (T, p) \in [\ -2.10 \mathrm{\ eV}, -1.75 \mathrm{\ eV} ]\ 
\end{equation}

%\bibliography{ref}

% %%%%%%%%%%%%%%% START: SI at the end of the paper #%%%%%%%%%%%%%%%%%%%%%%%%%%%%%%%%

% \onecolumngrid
% \setcounter{equation}{0}
% \setcounter{figure}{0}
% \setcounter{table}{0}
% \setcounter{page}{0}
% \makeatletter
% \renewcommand{\theequation}{S\arabic{equation}}
% % \renewcommand{\Figure\name}{{\bf Figure S}}
% \renewcommand{\tablename}{{\bf Table S}}
% \renewcommand{\figurename}{Fig. S\hspace{-1mm}}
% % \renewcommand{\tablename}{TABLE S\hspace{-1mm}}

% \include{SI}
% %%%%%%%%%%%%%%% END: SI at the end of the paper #%%%%%%%%%%%%%%%%%%%%%%%%%%%%%%%%


\end{document}